\documentclass[aps, prb, onecolumn, notitlepage, pamssymb,  amsmath, superscriptaddress,  showpacs, 12pt, tightenlines, nofootinbib] {revtex4-1}
\usepackage{graphicx}
\usepackage{bm}
\usepackage{fancyhdr}

\fancyheadoffset{1.5 cm}

\begin{document}
	
\title{Dynamics of domain walls in weak ferromagnets} 
\author{Zvezdin, A.K.}
\affiliation{Prokhorov General Physics Institute, Russian Academy of Sciences,  Moscow 119991}
	
	\begin{abstract}
		
		\noindent Pis$'$ma Zh. Exp. Teor. Fiz. \textbf{29}, No. 10, 605-610 (20 May 1979)
		
		\vspace{5 pt}
		
		It is shown that the total set of equations, which determines the dynamics of the
		domain bounds (DB) in a weak ferromagnet, has the same type of specific
		solution as the well-known Walker’s solution for ferromagnets. We calculated
		the functional dependence of the velocity of the DB on the magnetic field, which
		is described by the obtained solution. This function has a maximum at a finite
		field and a section of the negative differential mobility of the DB. According to
		the calculation, the maximum velocity $ c \approx 2 \times 10^6$ cm/sec in YFeO$_3$ is reached at
		$H_m \approx 4 \times 10^3$ Oe.
		
	\end{abstract} 	
	
	\maketitle

		
	\section{}
	The Landau-Lifshitz equations for a double sublattice weak ferromagnet can be
	represented in the form~\footnote[1]{For definiteness, we examine a crystal of rhombic symmetry.}
	\begin{equation} \label{1}
	\begin{aligned}
	\dot{\bm{M} }_i &= \gamma \left [ \bm{M}_i ,\frac{\delta W }{\delta \bm{M}_i}\right ] + \alpha \left [ \bm{M}_i,\dot{\bm{M}_i} \right ], \ i= 1,2 \\
	W &= \frac{a}{2}m^2 + \frac{b_1}{2}l_x^2 + \frac{b_3}{2}l_z^2 + d_1 m_z l_x - d_3 m_x l_z - \bm{m} \cdot \bm{H}+ A  \left ( \nabla \bm{l} \right )^2 + A' \left ( \nabla \bm{m} \right )^2 \\
	\bm{m } &= \frac{\bm{M}_1+\bm{M}_2}{2 M}, \ \bm{l } = \frac{\bm{M}_1-\bm{M}_2}{2 M}, \ \frac{\delta}{\delta q }\equiv \frac{\partial }{\partial q}-\nabla \frac{\partial }{\partial \nabla q}
	\end{aligned}
	\end{equation}
	
	To describe the dynamics of the domain bound, let us go over to the angular variables
	$\theta$, $\phi$, $\epsilon$, and $\beta$ in which ($\epsilon \ll 1$, $\beta \ll 1$):
	\begin{equation} \label{2}
	\begin{aligned}
	l_x &= \sin \theta \cos \phi, \ l_y = \sin \theta \sin \phi, \ l_z = \cos \phi, \ m_z = - \epsilon \sin \theta\\
	m_x &= \epsilon \cos \theta \cos \phi - \beta \sin \theta \sin \phi, \ m_y = \epsilon \cos\beta \sin\phi + \beta \sin \theta \cos \phi.
	\end{aligned}
	\end{equation}
	
	To write Eqs. \eqref{1} we use the Lagrange formalism in the variables $\theta$, $\phi$, $\epsilon$, and $\beta$. The Lagrange function $L$, the dissipative function $F$, and the corresponding Euler equations are
	\begin{equation} \label{3}
	L = \frac{M}{\gamma} \left [ \dot{\phi} \epsilon \sin \theta - \dot{\beta} \cos \theta \right ] - W(\theta,\psi,\epsilon,\beta) 
	\end{equation}
	\begin{equation} \label{4}
	F = \frac{\alpha M}{2 \gamma}\left [ \dot{\theta}^2 + \sin ^2 \theta \left ( \dot{\phi}^2 + \dot{\beta}^2 \right )  + \dot{\epsilon}^2 + 4 \epsilon \sin \theta\cos \theta \dot{\phi} \dot {
	\beta}\right ]
	\end{equation}
	\begin{equation} \label{5}
	\frac{\partial }{\partial t}\frac{\partial L}{\partial \dot{\theta}} = \frac{\delta L}{ \delta \theta} - \frac{\partial F}{\partial \dot{\theta}}, \text{ etc.}
	\end{equation}

	\section{} 
	 At $H = (0,0,H)$ we have a specific solution of the nonlinear equations \eqref{5}~\cite{walker}$^,$~\footnote[2]{It has the same meaning as the well-known Walker’s solution \cite{walker} for ferromagnets, although its equations are more complex.} :
	$\theta = \pi/2$, $\beta = 0$, $\phi(r,t)$, and $\epsilon(r,t)$. As a result of substituting this solution in Eqs. \eqref{5},
	two of the equations (obtained by varying $\theta$ and $\beta$) become identities and the other two have the form
	\begin{subequations}
	\begin{eqnarray}
	&\dot{\epsilon} + \alpha \dot {\phi} = \frac{c^2}{\omega_E}\nabla^2 \phi + \omega_1 \sin \phi \cos \phi - \omega_d\epsilon \sin \phi,\label{6a}
	\\
	&\alpha \dot {\epsilon} - \dot {\phi} =  \frac{{c'}^2}{\omega_E} \nabla^2 \epsilon - \omega_E \epsilon + \omega_d \cos \phi - \omega_H \label{6b}
	\end{eqnarray}
\end{subequations}
	
	\noindent where
	\begin{eqnarray*}
	\begin{aligned}
	\omega_1 &= \frac{\gamma b_1}{M}, \ \omega_d = \frac{\gamma d_1}{M}\equiv \gamma H_d, \ \omega_H = \gamma H, \\
	\omega_E &= \frac{\gamma a}{M} \equiv 2 \gamma H_E, \ c^2 = 4 \gamma^2 A H_E / M, \ {c'}^2 = 4 \gamma^2 A' H_E / M
	\end{aligned}
	\end{eqnarray*}
	
	First, let us determine the approximate solution of Eqs. \eqref{6a} and \eqref{6b}. In Eq. \eqref{6b} the terms $\left({c'}^2/\omega_E \right) \nabla^2 \epsilon$ and $\alpha \dot \epsilon$ can be deleted in comparison to $\omega_E$. The parameters of
	smallness of the deleted terms are $\left( a_0 / \Delta \right) ^2$ and $\left( \alpha a_0 / \Delta \right) ^2$, where
	$a_0 \left( c / \omega_E \right)  = \left(2 A / a \right ) ^ {1/2} = 10^{-8} $ cm and $\Delta$ is the thickness of the moving domain
	bound. Thus, we have from Eq. \eqref{6b}
	\begin{equation} \label{7}
	\epsilon = \frac{1}{\omega_E}\left ( - \omega_d \cos \phi + \omega_H + \dot{\phi} \right )
	\end{equation}

	Substituting it in Eq. \eqref{6a}, we obtain
	\begin{equation} \label{8}
	\ddot{\phi}-c^2 \nabla^2 \phi + \omega_A^2 \sin \phi \cos \phi = \dot{\omega}_H - \omega_d \omega_H \sin \phi - \alpha \omega_E \dot{\phi}
	\end{equation}
	
	where $\omega_A^2 = \omega_d^2 - \omega_E \omega_1$. At $H = 0$ and $\alpha = 0$ this equation becomes the well-known Sine-Gordon equation. Its one-dimensional solution, which satisfies the boundary con-
	ditions $\phi (x \rightarrow -\infty) = 0$, $\phi (x \rightarrow +\infty) = \pi$, has the form
	\begin{equation} \label{9}
	\phi(x,t) = 2 \arctan e^{\frac{x - v t}{\Delta}}, \ \Delta^{-1} = \frac{\omega_A/c}{\sqrt{1 - (v/c)^2}},
	\end{equation}

	where $v < c$. This function satisfies Eq. \eqref{8} at $H = const \neq 0$ and $\alpha \neq 0$, but for a specific	value of $v ( H )$, satisfies the equation, $\omega_d \omega_H = \alpha \omega_E \delta^{-1} v$, which can be easily verified
	by substituting Eq. \eqref{9} in Eq. \eqref{8}. From the last equation using Eq. \eqref{9} we obtain \cite{gyorgy1968analysis}$^,$\footnote[3]{	A similar dependence $v(H )$ was obtained in Ref. [\onlinecite{gyorgy1968analysis}], where the authors assumed that the dynamics of the weak ferromagnetic moment are described by the same equations as the dynamics of the ferromagnet and the magnetization remains constant during the motion of the domain bounds.}:
	\begin{equation} \label{10}
	v(H) = c \frac{H H_d}{a}\left [ 4 H_E^2 H_A^2 + a^{-2} H^2 H_d^2 \right ]^{-1/2}.
	\end{equation}
	
	The physical nature of such a dependence $v(H )$ can be described if we use the mechanical analogy of the motion of DB. If the dependence of $H$ and $v$ on $t$ is sufliciently small
	(the characterisitic frequencies of their variation are much smaller than $\omega_d$), we can
	obtain from Eq. \eqref{8} the following equation for the velocity of DB
	\begin{equation} \label{11}
	\frac{\mathrm{d} }{\mathrm{d} t}(mv) + \frac{(mv)}{r} = 2 M_s H,
	\end{equation}
	
	\noindent where
	\begin{equation*} 
	m = \frac{2 M_s}{H_d \gamma^2 \Delta(v)}=\frac{m_0}{\sqrt{1 - (v/c)^2}}, \ \tau = \frac{1}{\alpha \omega_E}.
	\end{equation*}

	All the terms in Eq. \eqref{11} have a clear mechanical meaning; $mv/\tau$ is the frictional force acting on the domain bounds, $2 M_s H$ is the pressure exerted on the domain bounds, etc. At $\left(d / d t \right) (m v) = 0 $ Eq. \eqref{11} gives Eq. \eqref{10}. Thus, the velocity of the domain 	bounds saturates as $H \rightarrow \infty$ because of the “relativistic” dependence of the mass $m$ of the DB on its velocity. Chetkin et al. \cite{chetkin1977velocity,chetkin1978maximum} observed experimentally and investigated the effect of saturation of the velocity of DB in YFeO$_3$; they \cite{chetkin1978maximum} as well as Bar’yakhtar et
	al. \cite{baryakhtar1978limit} theoretically estimated the limiting velocity of the DB in orthoferrites.

	\section{} 
	
	At $v \sim c$ the deleted terms in Eq. \eqref{6b} should be taken into account. Let us
	analyze asymptotically Eqs. \eqref{6a} and \eqref{6b} by the method proposed and developed in
	Refs.  [\onlinecite{schlomann1971structure}] and [\onlinecite{eleonskiy}]. Let us linearize Eqs. \eqref{6a} and \eqref{6b} near the stationary points $\phi = O$ and
	$\pi$, which correspond to the domains, and let us find solutions of the linear equations in
	the form $\mathrm{exp}\left[\pm \left(\omega_E/c \right) k (x - v_{\mp} t) \right] $ at $(x - v_{\mp}t) \rightarrow \mp \infty$. The conditions for the existence of nontrivial solutions have the form
	\begin{equation} \label{12}
	\left ( \frac{v}{c} \right )^2 \left ( 1 + a^2 \right ) - \frac{a v }{c} k\left [ 2 - k^{-2}\left ( 1 + \frac{\omega^2_A}{\omega^2_E} \mp \frac{\omega_H \omega_d}{\omega_E^2} \right ) \right ] - 1 + k^2 - \left ( 1 - k^{-2} \right ) \left ( \frac{\omega^2_A}{\omega^2_E} \mp \frac{\omega_H \omega_d}{\omega_E^2}  \right ) = 0.
	\end{equation}

	Let us assume that there is a solution of the nonlinear equations \eqref{6a} and \eqref{6b} in the form $\phi (x - vt )$, $\epsilon (x - vt)$ and that the function $\phi (x - vt )$ is symmetric; thus the equality $v_{+} (k) = v_{-} (- k) = v$, where $v_{+} (k)$ and $v_{-} ( k)$ are determined by Eq. \eqref{12}, gives (in the linear approximation of $\alpha$ and $H/H_E$):
	\begin{subequations}
		\label{13}
		\begin{eqnarray}
		v&=&c \left ( 1 + p - pk^{-2} - k^2  \right )^{1/2},\label{13a}
		\\
		H&=&H_{MC} k \left (1-p k^{-2}  \right )^{1/2} \left ( 1 - k^2 \right )^{-1/2} \left ( 1 + p - 2 k^2 \right ), \label{13b}
		\end{eqnarray}
	\end{subequations}

	\noindent	where
	\begin{equation*} 
	 H_{MC}  = a \frac{4 H_E^2}{H_d}, \ p = \left ( \frac{\omega_A}{\omega_E} \right )^2.
	\end{equation*}

		\begin{figure}[h!] 
		\includegraphics[width =0.6\columnwidth]{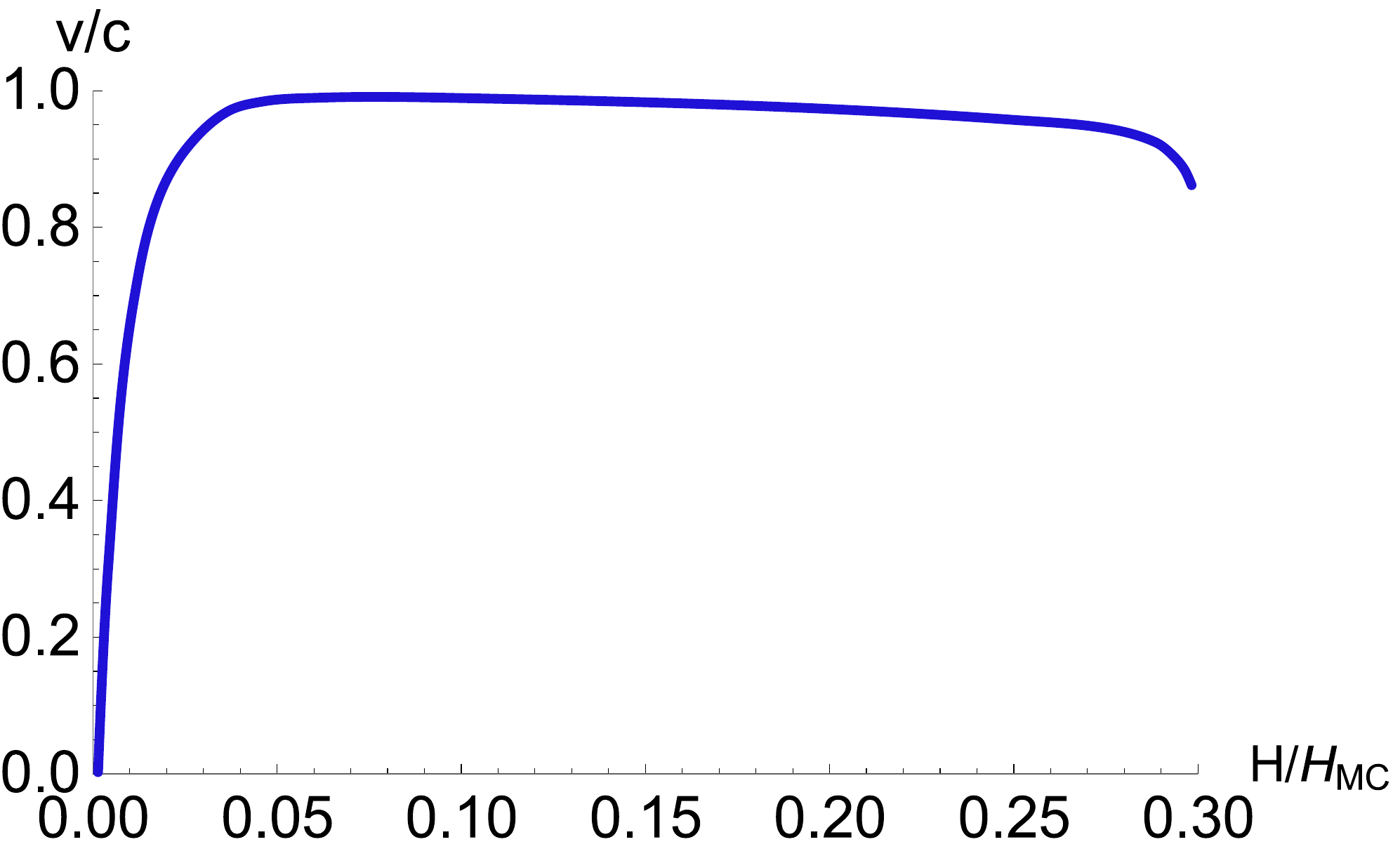} 
		\caption{A plot of the $v(H )$ function constructed according to Eq. \eqref{13} at $p = 10^4$, a part of the $v(H )$ curve
			$\left[ (H / H_{MC}) > 0.2\right]$ requires further study since here the condition $\epsilon \ll 1$ is violated.}\label{fig1}
	\end{figure}

	These equations determine the function $v(H )$ in the parametric form. The characteristic shape of this curve is given in Fig. \ref{fig1}. The maximum of this curve, which has the coordinates \footnote[4]{This velocity coincides with that obtained in Ref. [\onlinecite{baryakhtar1978limit}].}: $v_m = c(1 - \sqrt{p})$, $H_m =H_{MC}p^{1/4}\left( 1- \sqrt{p}\right) ^2$, corresponds to
	$k_m = p^{1/2} (p \ll 1)$ and the point $H = 0$, $v = 0$ corresponds to $k_m = p^{1/6} (p \ll 1)$. The quantity $k_m / k_0 \approx p^{1/4}$ characterizes the thickness ratio of the DB at $v = v_m$ and $v = 0$. The
	function \eqref{13} coincides with Eq. \eqref{10} at $k_0<k<k_m$ or at $0<H<H_m$. The last inequality is the condition of applicability of Eqs. \eqref{8} and \eqref{10}. The motion of the DB, in which $\bm l$ rotates in the $ac$ plane, is determined by more complicated equations than \eqref{6a} and
	\eqref{6b}, but the function $v(H )$, which is determined by Eqs. \eqref{10}, \eqref{13a}, and \eqref{13b}, remains valid in this case (if $d_1 = - d_3$). In them it must be assumed that $\omega_a = \left( b_1 - b_3\right)/M $.

	\section{} 
	
	We give the numerical estimates. In YFeO$_3$ $A \approx 4 \times  10^{-7}$ erg/cm, $H_E = 
	6.4 \times 10^6$ Oe, $H_d = 10^5$ Oe, and $p \approx 10^{-4}$. The “scale” of the field H MC can be expressed in terms of the mobility $\mu$ when $H \rightarrow 0$: $\mu = \left( c / H_{MC} \right) p^{1/2}$. Hence, $H_{MC} = \left( c \sqrt{p} / \mu\right) $. According to Ref.  [\onlinecite{uait}], $\mu \simeq  5 \times 10^3 $ cm/sec$\cdot$Oe. Using these values, we obtain $c \approx 2 \times 10^6$ cm/sec, $H_{MC} \approx 4 \times 10^4$ Oe, $V_m = 0.99$ s, and $H_m\approx 4 \times 10^3 $ Oe.

\end{document}